# Decoding enzyme–substrate interaction topology reveals principles underlying catalytic efficiency and mutational outcomes

Weiren Zhao[1,2], Takeyuki Tamura[*,1].

*[1] Bioinformatics Center, Institute for Chemical Research, Kyoto University*

*2 Department of Intelligence Science and Technology, Graduate School of Informatics, Kyoto University*

*[*]Corresponding author: Takeyuki Tamura (tamura@kuicr.kyoto-u.ac.jp)*

## Abstract

The enzyme turnover number ($k_{cat}$) defines catalytic efficiency and constrains quantitative models of metabolism, yet the molecular determinants governing $k_{cat}$ and its response to mutation remain poorly understood. Measurements are sparse and labor-intensive, and most computational approaches provide numerical predictions without explaining how enzyme–substrate interactions shape catalytic outcomes. A central challenge is therefore to identify the topological principles that determine where mutations act and how their functional outcomes are encoded within the enzyme–substrate interaction network.

Here, we show that catalytic efficiency and mutational effects can be interpreted through enzyme-substrate interaction topology. We developed Interkcat, an interpretable bidirectional cross-attention framework that captures reciprocal coordination between protein residues and substrate atoms. Optimized on a unified benchmark, Interkcat achieves state-of-the-art predictive performance ($R^2$ = 0.701). From its learned representations, we derive an Interaction Topology Score (ITS) that identifies sequence regions statistically enriched for mutation-sensitive sites without explicit structural inputs. We further demonstrate that higher-order topological features distinguish opposing mutational outcomes: lethal mutations disrupt coordinated networks, whereas activity-preserving or enhancing mutations retain sparse, globally organized coupling.

These findings establish interaction topology as a unifying principle linking enzyme sequence, catalytic efficiency, and evolutionary perturbation.

## Introduction

The turnover number ($k_{cat}$) is a fundamental kinetic parameter that defines the maximal rate at which a single enzyme active site converts substrates into products[1]. It serves as a key parameter for the quantitative understanding of cellular metabolism, physiology, and resource allocation[2-6]. Specifically, accurate genome-scale values are indispensable for constructing enzyme-constrained genome-scale metabolic models (ecGEMs)[7, 8]. By integrating these kinetic constraints, ecGEMs significantly enhance the accuracy of simulating metabolic shifts, maximal growth rates, and proteome allocation patterns across diverse organisms[9, 10].

Despite its critical importance, the experimental determination of $k_{cat}$ remains labor-intensive, expensive, and low-throughput[11, 12]. Consequently, the landscape of available kinetic data is remarkably sparse[13]. Even in the biochemically best-characterized model organism, Escherichia coli, in vitro values are documented for only ~10% of catalyzed reactions[14]. This data gap is even more pronounced in other organisms; for instance, the coverage in Saccharomyces cerevisiae ecGEMs is merely ~5%[15, 16]. Furthermore, existing experimental data are often noisy, exhibiting substantial variance due to inconsistent assay conditions such as pH, temperature, and cofactor availability.

Deep learning (DL) models have emerged as a transformative solution to bridge this data gap, enabling large-scale, cost-effective predictions of enzyme efficiency[17, 18]. Recent advances, such as TurNuP[1], DLKcat[10] and CatPred[17], have successfully utilized protein sequences and substrate structures to predict $k_{cat}$ with high throughput. However, most previous computational methods have focused primarily on numerical regression, often operating as "black-box" predictors that lack transparency regarding the underlying molecular mechanisms. While some models have attempted to incorporate attention mechanisms to identify key residues, they often struggle to generalize across diverse biological contexts or provide granular, actionable insights for enzyme engineering[18].

In this work, we introduce Interkcat, an interpretable bidirectional cross-attention framework designed for precise enzyme catalytic efficiency prediction. By integrating ESM-2 and Graph Neural Networks (GNN), we formulated the Interaction Topology Score (ITS), an attention-derived metric that integrates molecule-to-protein (m2p) targeting intensity with protein-to-molecule (p2m) interaction specificity. This design transitions deep learning from a traditional regressor to a topology-aware discovery tool. By validating the model against mutation datasets, we demonstrate its capacity to identify sequence regions statistically enriched for mutation-sensitive sites, all without requiring explicit 3D structural inputs.

The practical significance of Interkcat extends beyond numerical estimation to rational enzyme engineering and directed evolution[21-26]. Beyond locating sensitive functional cores, the framework reveals that directional mutational outcomes, whether lethal or active, are encoded in distinct topological signatures of the interaction network. We show that features such as Peak Count, SVD Dominance, and Global Gini indices successfully differentiate these effects, providing a strategic roadmap for engineering efforts. This approach allows researchers to safeguard essential catalytic motifs while prioritizing high-plasticity residues for optimization, thereby accelerating the development of industrial biocatalysts with superior turnover efficiencies and improved metabolic performance.

# Results

## Establishment of a Unified Benchmark for Fair Comparison

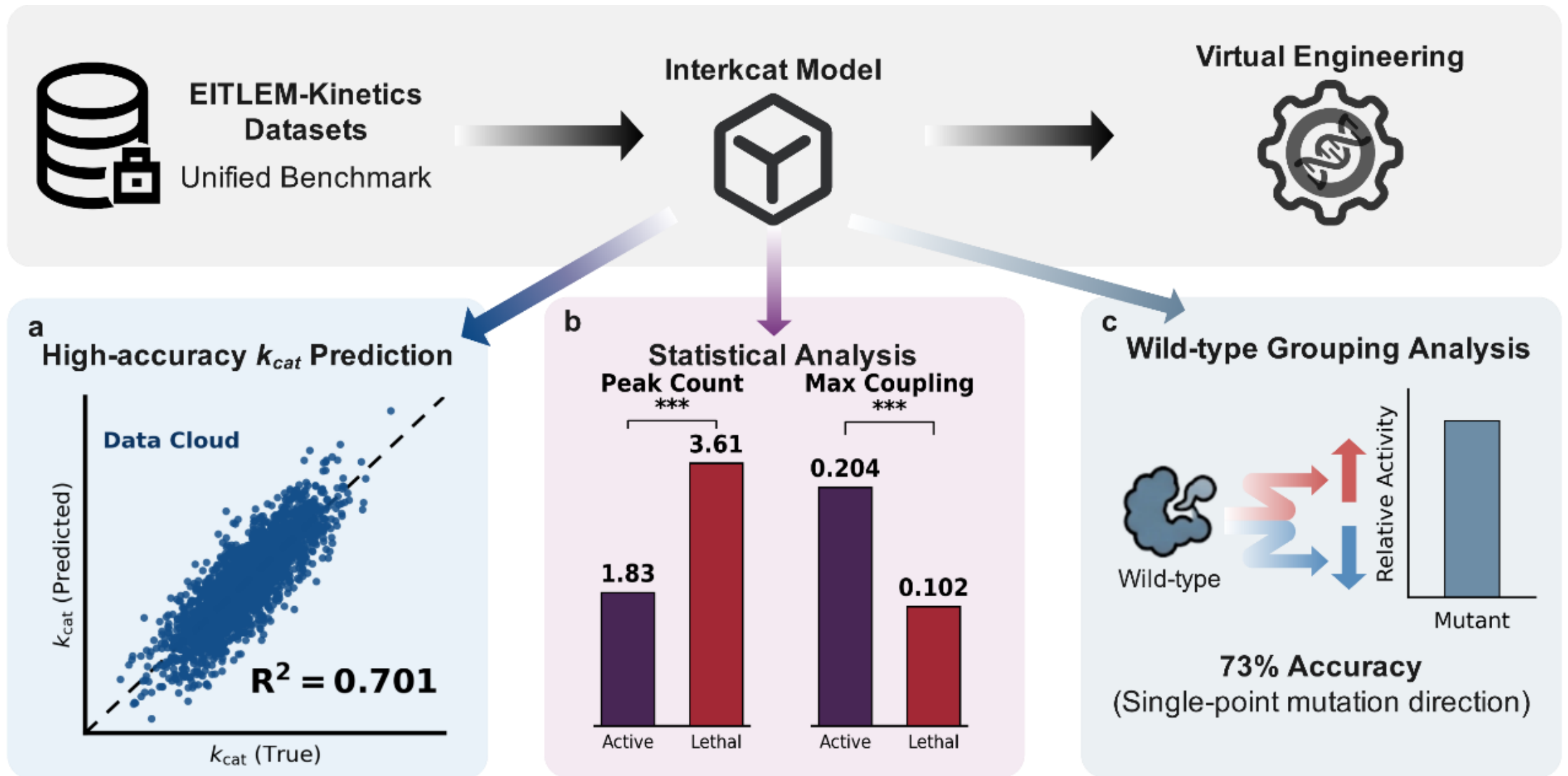

**Fig. 1 | Overview of the Interkcat framework for enzyme kinetics prediction, mechanistic statistical analysis, and virtual engineering.** The Interkcat model leverages the unified EITLEM-Kinetics benchmark dataset to decode enzyme-substrate interactions, establishing an end-to-end pipeline for rational virtual enzyme design. **a** High-accuracy $k_{cat}$ prediction. The model achieves state-of-the-art predictive performance for continuous turnover numbers, demonstrating a robust correlation with true experimental values ($R^2 = 0.701$). **b** Statistical differentiation of mutation effects. Quantitative analysis reveals highly significant mechanistic disparities between active and lethal mutations (*** $p < 0.001$). Lethal mutations are characterized by a higher Peak Count (mean 3.61) compared to functional variants (1.83). Conversely, active mutations exhibit a significantly higher Max Coupling score (0.204) than their lethal counterparts (0.102), highlighting the structural and evolutionary constraints required for maintaining catalytic activity. **c** Application in wild-type grouping. By evaluating single-point mutations within distinct wild-type groups, the model accurately predicts the directional shift in enzyme activity (enhancement or reduction) with an overall accuracy of 73%. **Source data are provided as a Source Data file.**

Previous computational studies on enzyme kinetics have typically relied on independently collected and custom-cleaned datasets, creating a fragmented landscape that precludes fair, head-to-head comparisons of model performance. To overcome this pervasive challenge and rigorously evaluate the Interkcat framework, we adopted the unified EITLEM-Kinetics[27] dataset as our standardized training and evaluation corpus. Detailed information regarding the EITLEM-Kinetics dataset is provided in the Supplementary Information (Supplementary Fig. 1).

Recently established through comprehensive benchmarking efforts by Chen, Gao, and colleagues[13], EITLEM-Kinetics serves as a robust standard for the field. It comprehensively integrates curated data from major repositories, namely BRENDA[28], SABIO-RK[29] and UniProt[30], representing the largest and highest-quality continuous kinetic collection available to date. Comprising 34,140 curated data points, it supports state-of-the-art predictive baselines with reported test $R^2$ values reaching up to 0.667.

By utilizing this unified benchmark, we ensure an equitable evaluation of Interkcat's continuous predictive capabilities against existing models, while simultaneously establishing a statistically rigorous foundation for our subsequent mechanistic interpretability and virtual engineering analyses. Specifically, our research framework is structured around three core components: training the designed Interkcat model for high-accuracy prediction (Fig. 1a), employing rigorous statistical methods for mechanistic model interpretation (Fig. 1b), and exploring the practical application of the trained model in predicting directional shifts in enzyme activity (Fig. 1c).

## Architecture of the Interkcat Framework

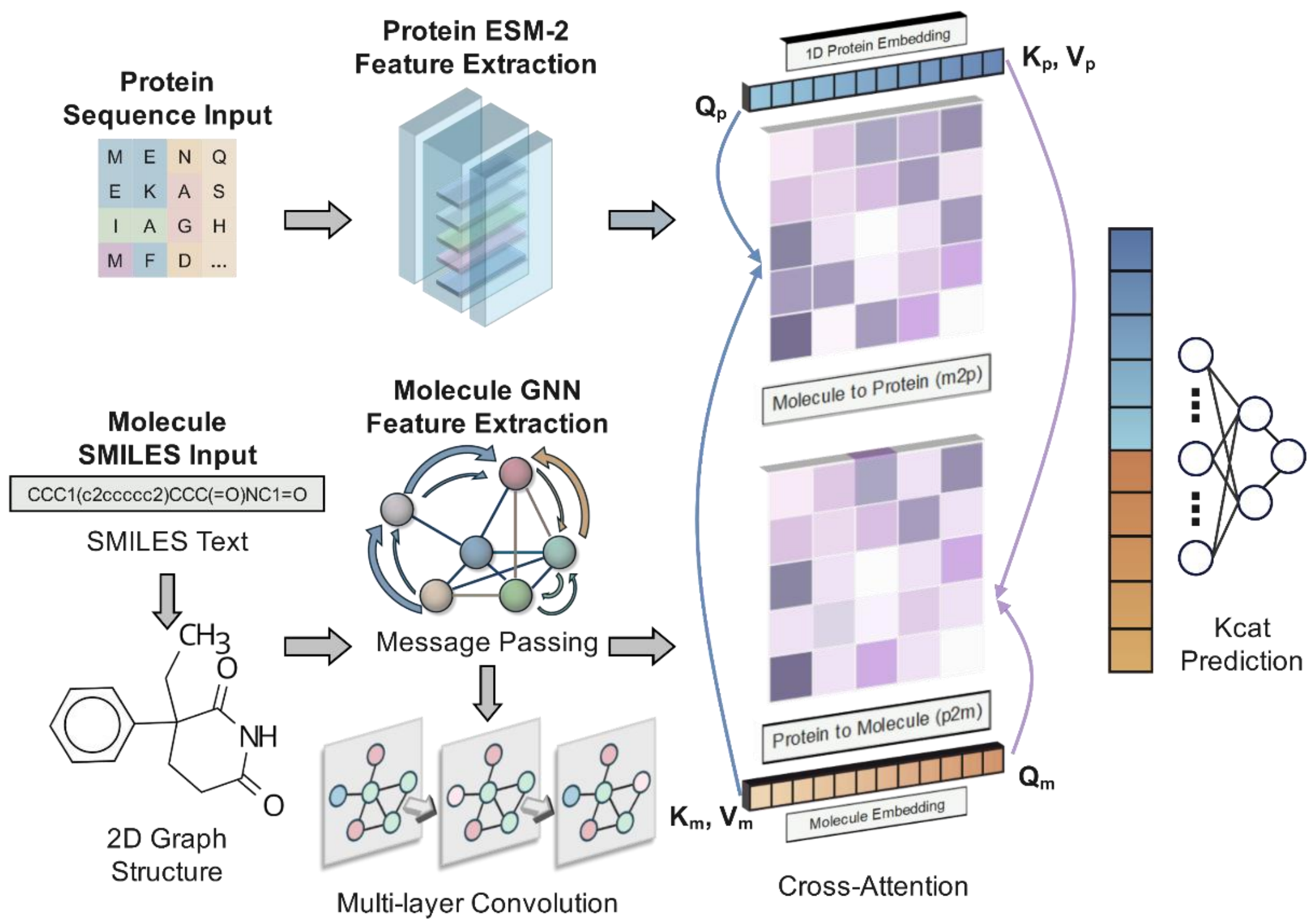

**Fig. 2 | Schematic illustration of the Interkcat model architecture for enzyme turnover number ($k_{cat}$) prediction.** The framework integrates a pre-trained ESM-2 module for protein sequence feature extraction and a GNN for 2D molecular graph representation. The core interaction between the enzyme and substrate is captured via a bidirectional cross-attention mechanism, comprising Molecule-to-Protein and Protein-to-Molecule information flows driven by respective Query (*Q*), Key (*K*), and Value (*V*) matrices. The resulting inter-coordinated embeddings are concatenated and decoded by a multi-layer perceptron to predict the target $k_{cat}$ value.

Following the establishment of our evaluation framework, we detail the core architecture of the Interkcat model designed to decode complex enzyme-substrate interactions (Fig. 2). Traditional computational models often struggle to capture the granular, reciprocal interactions between a biocatalyst and its target molecule. To address this limitation, Interkcat is structured as an end-to-end deep learning pipeline utilizing a dual-branch feature extraction strategy integrated with a bidirectional cross-attention mechanism.

As illustrated in Fig. 2, linear protein sequences and substrate SMILES strings serve as the primary inputs. For the protein branch, we employ ESM-2, a pre-trained protein language model, to extract rich 1D protein embeddings that capture evolutionary and structural contexts. Concurrently, the substrate SMILES strings are converted into 2D molecular graphs, where nodes represent atoms and edges represent chemical bonds. These graphs are subsequently processed through a tailored GNN to generate comprehensive 1D molecular embeddings. The hallmark of the Interkcat framework lies in its ability to bridge these heterogeneous modalities. The extracted protein and molecular embeddings are projected into a shared latent space and utilized as Query, Key, and Value matrices. This configuration implements a bidirectional cross-attention mechanism (See Methods for more details), establishing two distinct information flows, Molecule-to-Protein (m2p) and Protein-to-Molecule (p2m), that enable the heterogeneous modalities to mutually contextualize their reciprocal structural and functional interplay. By computing these mutual attention weights, the model explicitly captures the reciprocal structural and functional interplay between the enzyme and the substrate. Finally, these inter-coordinated embeddings are concatenated and decoded by a fully connected network to execute the accurate continuous regression task.

## Model Training and Predictive Performance

**Table 1 | Predictive performance of Interkcat against existing computational models for continuous $k_{cat}$ prediction.**

| Models | Test Performance | | |
|---|---|---|---|
| | $R^2$ | *RMSE* | *MAE* |
| TurNuP[1] | 0.609 | 0.944 | 0.657 |
| DLKcat[10] | 0.548 | 1.057 | 0.734 |
| DLTKcat[18] | 0.498 | 1.092 | 0.778 |
| EITLEM-Kinetics ($k_{cat}$)[27] | 0.628 | 0.927 | 0.582 |
| UniKP ($k_{cat}$)[31] | 0.674 | 0.871 | 0.590 |
| DeepEnzyme[32] | 0.516 | 1.029 | 0.706 |
| CataPro ($k_{cat}$)[33] | 0.535 | 1.010 | 0.708 |
| **Interkcat** | **0.701** | **0.826** | **0.545** |

Performance metrics for all models were derived using the unified EITLEM-Kinetics dataset, partitioned into training (80%), validation (10%), and independent test (10%) sets using a standardized random seed (seed = 1234). Training was conducted for a maximum of 200 epochs with a learning rate of $1\times10^{-4}$. To ensure optimal generalization and preclude overfitting, an early stopping strategy with a patience of 20 epochs was implemented, whereby training dynamically halted upon reaching a performance plateau on the validation set. Standardized evaluation conditions were maintained across all baseline models to guarantee an equitable comparison.

Following the architectural formulation of the bidirectional cross-attention mechanism, we systematically evaluated the predictive capabilities of the Interkcat framework. To ensure a scientifically rigorous assessment, we benchmarked Interkcat against existing state-of-the-art models using a standardized evaluation protocol on the unified EITLEM-Kinetics dataset. This comparative analysis focuses on the model's ability to execute precise continuous regression while maintaining robust generalization across diverse enzymatic systems (Table 1). Detailed optimization trajectories and hyperparameter diagnostics are provided in Supplementary Fig. 2 and Supplementary Tables 1-5.

Beyond its predictive performance, Interkcat provides mechanistic transparency by decoding the internal interaction topology of enzyme-substrate pairs. Systematic investigation of the bidirectional cross-attention matrices reveals a fundamental topological asymmetry between the m2p and p2m information flows (Supplementary Figs. 3-8). Specifically, the m2p attention matrices exhibit a highly sparse and localized topology, where individual substrate atoms selectively target a restricted subset of amino acids likely corresponding to functionally sensitive regions. Conversely, the p2m matrices demonstrate a broader, more globally distributed pattern, suggesting that the enzyme backbone acts collectively to provide a coordinated structural scaffold for the substrate. To integrate the complementary information from both asymmetrical attention weights, we developed a metric termed the Interaction Topology Score (See Methods for computational details). Unlike previous computational studies that mainly report numerical predictions, this score provides a sequence-level topology profile for testing whether high-scoring regions are enriched for mutation-sensitive sites.

### Wild-type-Anchored Grouping Strategy for Comparative Mutation Analysis

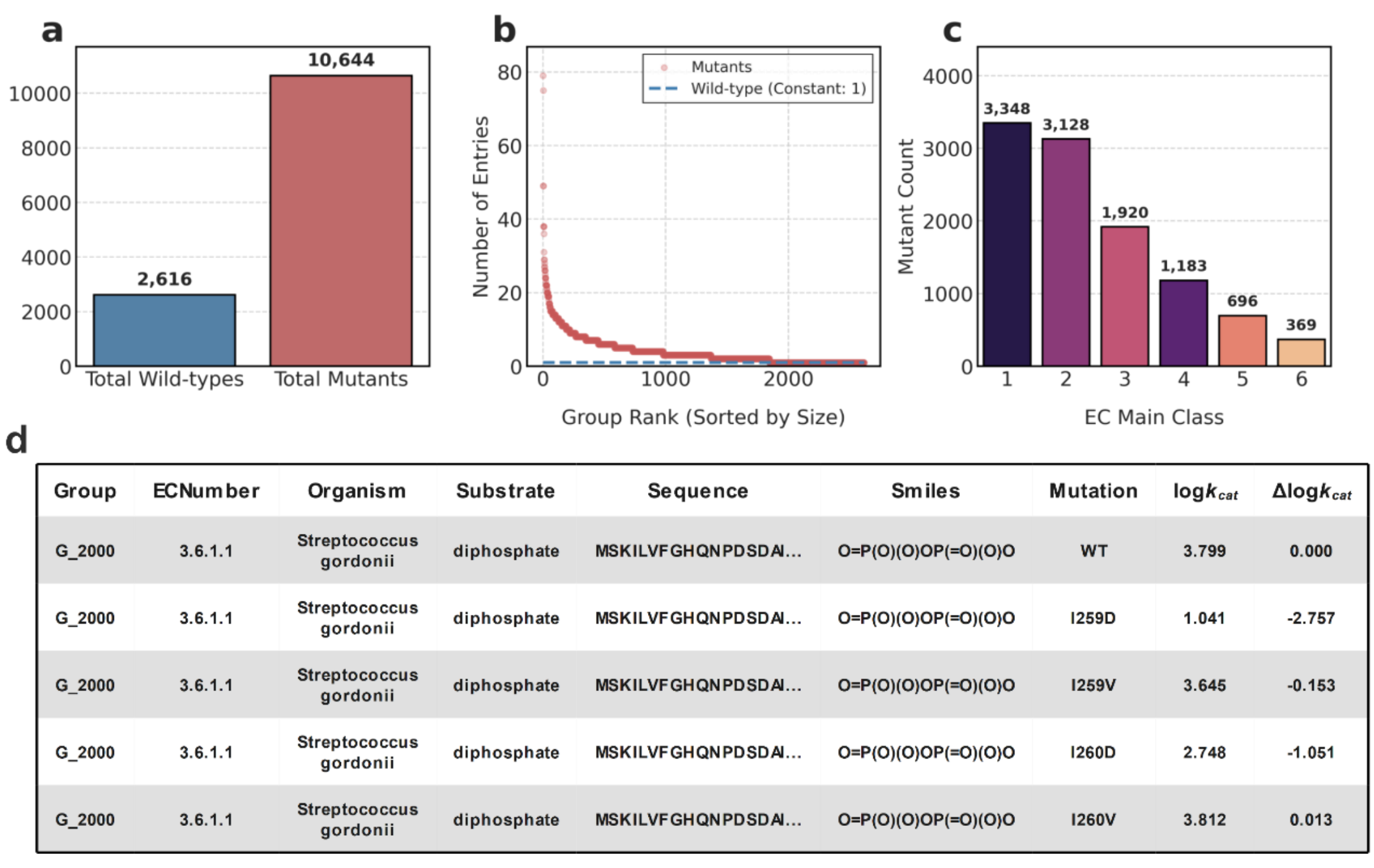

| Group | ECNumber | Organism | Substrate | Sequence | Smiles | Mutation | $\log k_{cat}$ | $\Delta \log k_{cat}$ |
|---|---|---|---|---|---|---|---|---|
| G_2000 | 3.6.1.1 | Streptococcus gordonii | diphosphate | MSKILVFGHQNPDSDAI... | O=P(O)(O)OP(=O)(O)O | WT | 3.799 | 0.000 |
| G_2000 | 3.6.1.1 | Streptococcus gordonii | diphosphate | MSKILVFGHQNPDSDAI... | O=P(O)(O)OP(=O)(O)O | I259D | 1.041 | -2.757 |
| G_2000 | 3.6.1.1 | Streptococcus gordonii | diphosphate | MSKILVFGHQNPDSDAI... | O=P(O)(O)OP(=O)(O)O | I259V | 3.645 | -0.153 |
| G_2000 | 3.6.1.1 | Streptococcus gordonii | diphosphate | MSKILVFGHQNPDSDAI... | O=P(O)(O)OP(=O)(O)O | I260D | 2.748 | -1.051 |
| G_2000 | 3.6.1.1 | Streptococcus gordonii | diphosphate | MSKILVFGHQNPDSDAI... | O=P(O)(O)OP(=O)(O)O | I260V | 3.812 | 0.013 |

**Fig. 3 | Dataset grouping strategy and distributional characteristics for wild-type and mutant**

**comparative analysis. a** Total absolute counts of wild-type enzymes and their corresponding mutant variants within the unified kinetic dataset after the grouping procedure. **b** Rank-ordered distribution of mutant entries per group. The data exhibits a long-tail distribution, with each group strictly anchored by a single wild-type reference (constant dashed line at $y = 1$). **c** Stratification of the mutant dataset across the six primary Enzyme Commission (EC) classes, illustrating the broad enzymatic diversity captured in the curated data. **d** A representative tabular snapshot of a designated group (Group G_2000). The panel details the alignment of the wild-type baseline and its associated point mutants alongside essential metadata (organism, truncated sequence, substrate SMILES). Crucially, it defines both the absolute catalytic activity (log $k_{cat}$) and the relative activity shift ($\Delta$log $k_{cat}$), establishing the foundational data structure for subsequent downstream directional prediction tasks. **Source data are provided as a Source Data file.**

To systematically leverage these ITS profiles for a comparative analysis of wild-type (WT) and mutant enzymes, we established a rigorous dataset grouping strategy within the EITLEM-Kinetics benchmark (Fig. 3). To effectively decouple true mutational effects from confounding factors, such as phylogenetic divergence or substrate variations, data points were partitioned into distinct cohorts sharing four identical contextual attributes: EC number, source organism, substrate name, and substrate SMILES. Within this framework, we strictly retained only those groups anchored by a single WT reference sequence accompanied by at least one corresponding mutant.

Furthermore, to evaluate functional shifts, mutants within each group were quantitatively categorized based on their relative activity changes ($\Delta\log_{10}$ $k_{cat}$). Specifically, we defined 'active' mutations as those exhibiting an activity increase of at least one order of magnitude ($\Delta\log_{10}$ $k_{\mathrm{cat}} \geq 1$), and 'lethal' mutations as those exhibiting an activity decrease of at least one order of magnitude ($\Delta\log_{10}\ k_{\text{cat}} \leq -1$). This threshold ($|\Delta\log_{10}\ k_{\text{cat}}| \geq 1$) was used to select mutants with substantial functional shifts for subsequent structural comparisons. Fig. 3 visually summarizes the outcomes of this grouping strategy. The distributional characteristics of the grouped dataset, including the absolute counts of WT and mutant entries, the long-tail distribution of mutant frequencies, and the stratification across primary EC classes, are comprehensively illustrated (Fig. 3a–c). Finally, Fig. 3d provides a representative snapshot of a designated group (Group G_2000), demonstrating the precise intra-group alignment of the WT baseline with its respective point mutants and their corresponding $\Delta\log_{10}\ k_{cat}$ shifts, thereby laying the essential data architecture for our subsequent virtual engineering analyses.

## Decoding Enzyme–Substrate Interaction Topology as a Locator for Functional Mutation Hotspots

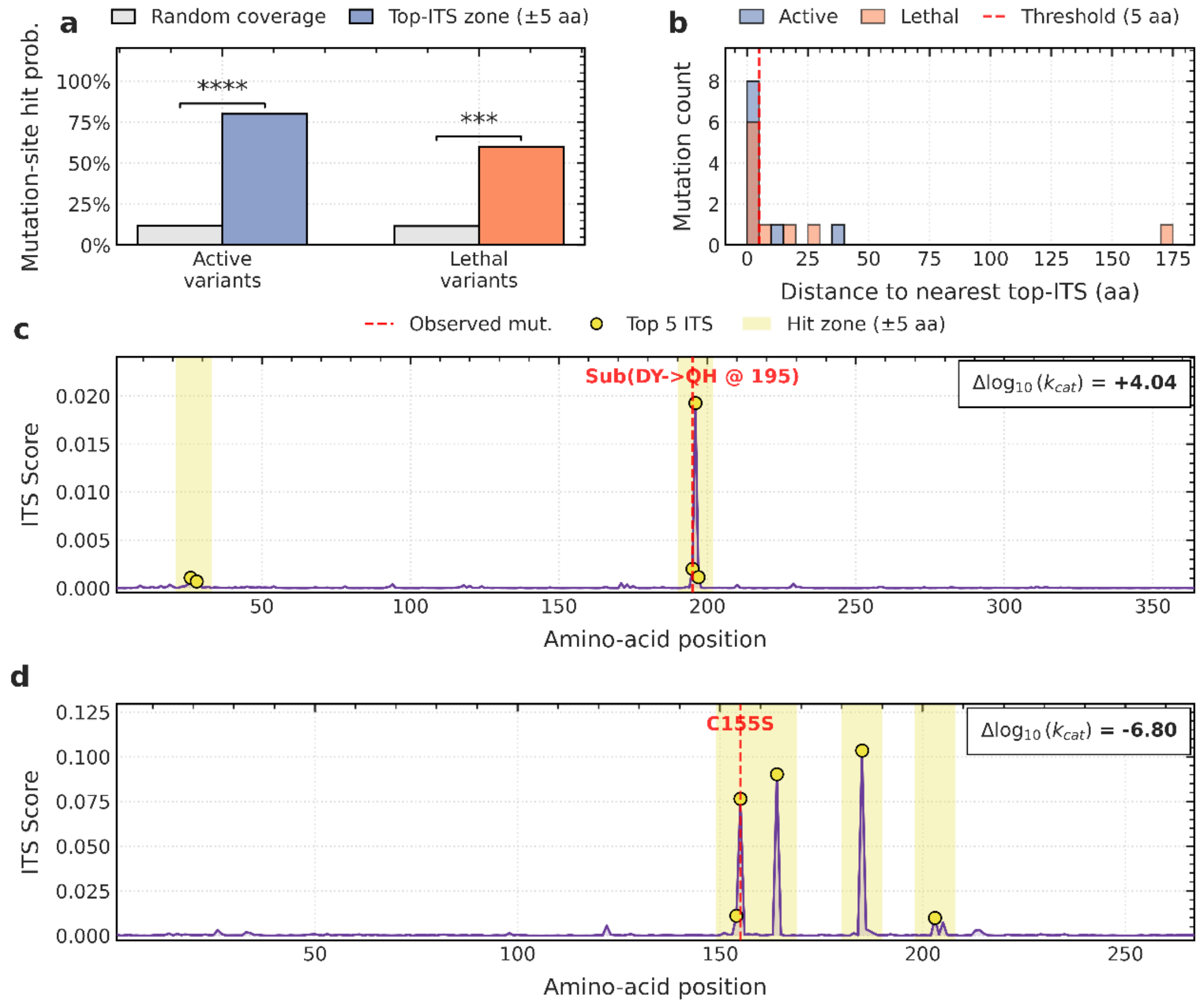

**Fig. 4 | Enzyme-substrate interaction topology identifies sequence regions enriched for high-shift mutation sites without structural inputs. a** Comparison of mutation site hit probabilities between the baseline random selection algorithm and the Interkcat-driven Top-ITS zone identification algorithm, evaluated across the top 10 representative samples from the active and lethal mutation cohorts. The Top-ITS zone algorithm defines consensus windows centered around local mathematical maxima of the interaction landscape without explicit structural inputs. **b** Spatial quantification of mutational distribution within the top 10 sample cohorts. The bar chart (left axis) details the absolute number of active and lethal mutation sites successfully localized inside the predicted hit zones. The dot plot (right axis) illustrates the tight sequence distance (in number of amino acid residues) from

unhit mutation sites to the nearest boundary of the predicted Top-ITS zone. **c, d** Representative sequence-level ITS profiles for an active variant (c) and a lethal variant (d). Shaded areas highlight the algorithmically prioritized Top-ITS zones. Vertical dashed lines mark the exact positions of the amino acid mutations. The inset in the top-right corner of each panel indicates the experimental catalytic efficiency shift ($\Delta \log_{10} k_{\mathrm{cat}}$) of the mutant relative to its corresponding wild-type (WT) anchor. **Source data are provided as a Source Data file.**

To systematically evaluate whether the spatial-functional landscapes uncovered by our wild-type-anchored grouping strategy could act as an actionable predictive compass, we investigated the direct correspondence between the topological profiles and physical mutation sites (Fig. 4). We hypothesized that the bidirectional cross-attention landscapes could provide attention-derived topology profiles that identify sequence regions enriched for mutations associated with large catalytic shifts.

We selected the ten most representative variants from each of the activity-enhancing and lethal mutation cohorts for topological validation (Fig. 4a). We then compared the empirical frequency of observed mutation sites within algorithmically defined Top-ITS zones (See Methods for more details) with that expected from random sequence windows. Top-ITS zones captured mutation sites far more often than the random baseline, indicating that functional mutations are not uniformly distributed along the sequence but concentrate within defined interaction-topology regions. Individual profiles are shown in Supplementary Figs. 12-33.

The two cohorts showed distinct but related spatial patterns (Fig. 4b). Mutations falling inside Top-ITS zones clustered strongly in both activity-enhancing and lethal variants, placing gain and loss of function within the same high-scoring topological regions. For mutations outside these zones, we measured the linear sequence distance to the nearest Top-ITS boundary. These out-of-zone mutations remained close to the predicted zones, consistent with local perturbation of residues surrounding the catalytic core.

We next mapped ITS profiles onto measured functional changes for representative variants (Fig. 4c, d). In the activity-enhancing variant, engineered substitutions coincided with major ITS peaks and with increased catalytic activity relative to wild type ($\Delta \log_{10} k_{\mathrm{cat}} > 0$). In the lethal variant, substitutions at critical ITS peaks coincided with a strong loss of turnover ($\Delta \log_{10} k_{\mathrm{cat}} \ll 0$). Together, these analyses show that Top-ITS zones identify sequence regions statistically enriched for mutation-sensitive sites.

## Statistical differentiation of mutational effects

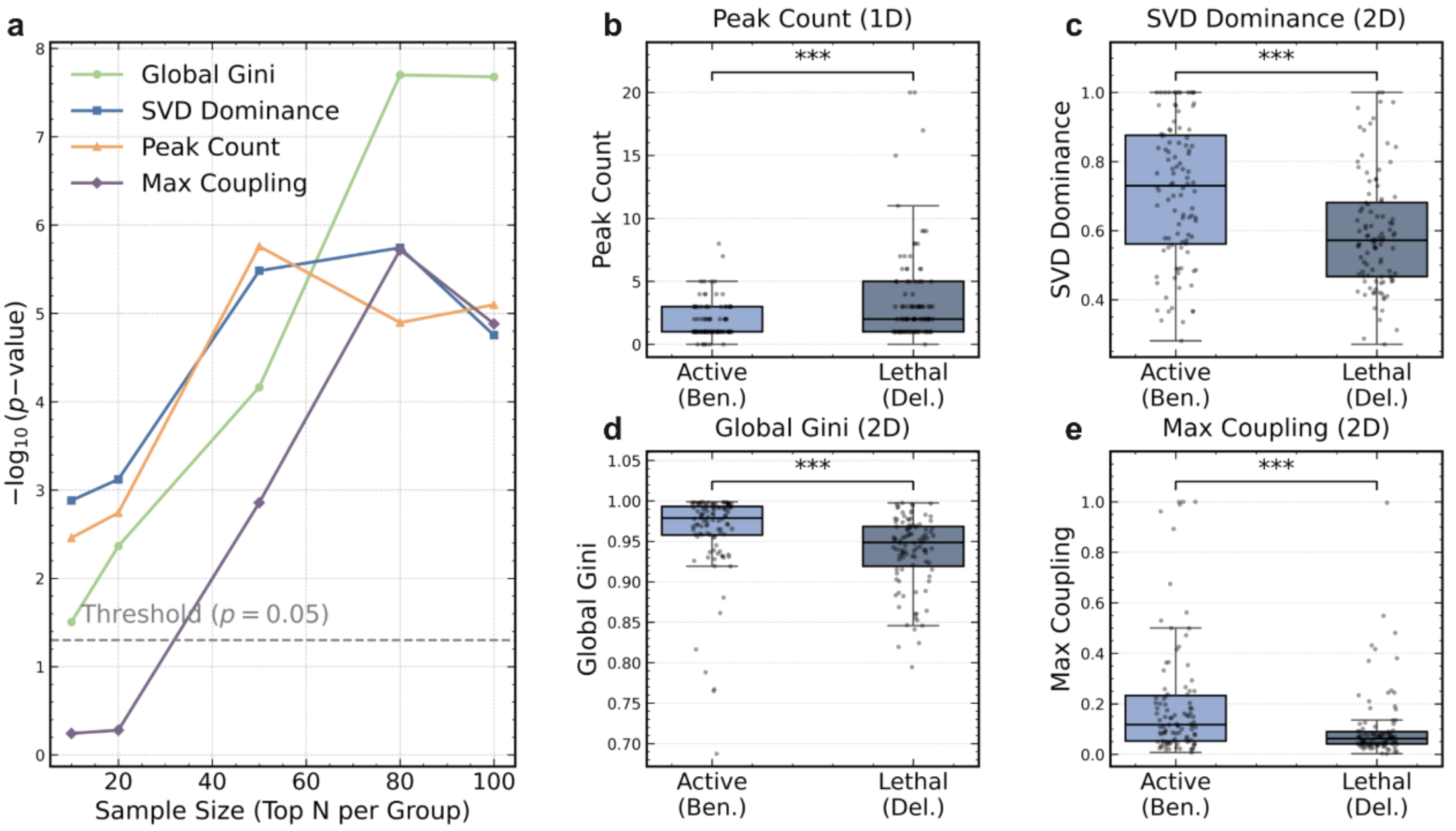

**Fig. 5 | Discriminative power of attention and entropy metrics for mutation effects. a** Statistical significance trajectory of four derived features across varying sample sizes, confirming robust separation between mutation groups well above the P = 0.05 threshold (dashed line). **b–e** Direct

statistical comparisons between active and lethal mutations. Lethal variants are characterized by a significantly higher 1D Peak Count (**b**). In contrast, functionally active mutations maintain significantly higher 2D structural coordination, as evidenced by elevated SVD Dominance (**c**), Global Gini (**d**), and Max Coupling (**e**) scores. Statistical significance between the active mutation group (n=100) and the lethal mutation group (n=100) across various topological features was determined using the non-parametric two-sided Mann-Whitney U test. As a non-parametric method, this test does not assume a specific underlying probability distribution for the data. Instead, it only assumes that the observations are independent and the variables are continuous, evaluating the probability that a randomly selected value from one group is greater or less than a randomly selected value from the other. P values are indicated as follows: * P < 0.05, ** P < 0.01, *** P < 0.001, **** P < 0.0001. **Source data are provided as a Source Data file.**

Using the standardized wild-type-anchored mutation groups, we next asked whether activity-enhancing and lethal mutations differed in higher-order interaction topology. Although 1D positional metrics such as ITS identified enriched mutation-sensitive regions, they did not by themselves resolve the direction of mutational effects. We therefore extracted four mechanistic features from the bidirectional attention matrices: 1D Peak Count and three 2D coordination metrics, SVD Dominance, Global Gini index and Max Coupling (See Methods for more details).

Across increasing sample sizes from 10 to 100, the statistical separation between lethal and activity-enhancing mutations increased consistently for these features, as measured by $-\log_{10} p$ (Fig. 5a)[34, 35]. We used the non-parametric Mann-Whitney U test to compare feature distributions without assuming a specific underlying distribution. All four features exceeded the significance threshold ($p < 0.05$), indicating reproducible differences between the two mutation classes. At a representative sample size of 100, direct feature comparisons resolved the spatial basis of this separation (Fig. 5b-e). Lethal variants showed a higher 1D Peak Count (Fig. 5b), consistent with fragmented perturbations across the catalytic region. By contrast, activity-enhancing variants retained stronger 2D coordination, with higher SVD Dominance, Global Gini index and Max Coupling scores (all $p < 0.001$; Fig. 5c-e). Exact mean values and statistical results are provided in Supplementary Table 6.

Together, these results indicate that lethal mutations disrupt the enzyme-substrate interaction network, whereas activity-enhancing mutations preserve coordinated topological constraints compatible with catalysis. Thus, moving from 1D hotspot localization to multidimensional topology distinguishes where mutations occur from how they reshape catalytic function.

## Practical Application in Virtual Enzyme Engineering

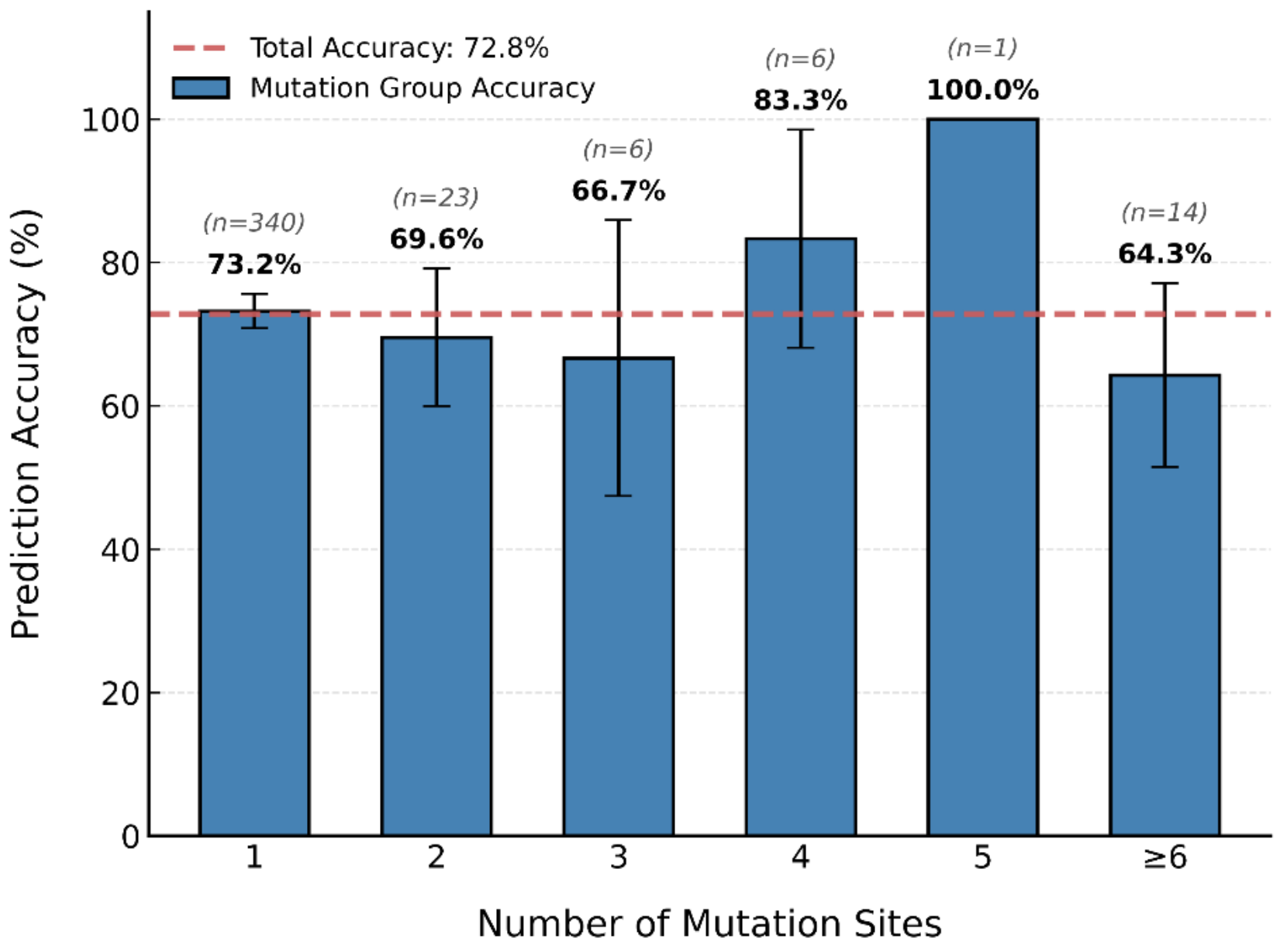

**Fig. 6 | Generalization performance of the Interkcat model in predicting mutation effects on unseen data.** By applying the established wild-type-anchored grouping strategy to this unseen subset, a final curated evaluation cohort of 390 mutant entries ($N = 390$) was obtained. The bar chart stratifies

the prediction accuracy, defined as the correct classification of the directional shift in enzyme activity, by the number of simultaneous mutation sites per variant. The horizontal red dashed line highlights the robust overall prediction accuracy of 72.8% across the entire evaluation cohort. Notably, the model demonstrates high reliability for single-point mutations (73.2% accuracy, $n = 340$ ) while maintaining competitive predictive power across more complex, multi-point mutant configurations. **Source data are provided as a Source Data file.**

Finally, to demonstrate the practical utility of Interkcat for rational enzyme design, we evaluated its generalization performance in predicting mutational effects on completely unseen data (Fig. 6). Applying our established wild-type-anchored grouping strategy, we compiled a curated evaluation cohort of 390 independent mutant entries. The model successfully predicted the directional shifts in enzyme activity with a robust overall accuracy of 72.8%. Notably, Interkcat exhibited strong reliability in evaluating single-point mutations, achieving an accuracy of 73.2% ( $n = 340$ ), while maintaining competitive predictive power for more complex, multi-point variants.

## Discussion

This study identifies enzyme-substrate interaction topology as a mechanistic layer linking catalytic efficiency with mutational outcome. Using a unified kinetic benchmark, Interkcat achieved improved continuous $k_{cat}$ prediction while exposing bidirectional attention patterns between protein residues and substrate atoms. The central finding is not only that $k_{cat}$ can be predicted more accurately, but that the learned interaction topology contains interpretable information about where functional mutations occur and how their effects diverge.

The ITS analysis suggests that mutation-sensitive residues are not distributed randomly across enzyme sequences. Instead, activity-enhancing and lethal mutations were enriched near Top-ITS zones, indicating that both beneficial and disruptive substitutions often act through the same interaction-topology landscape. This observation supports a topology-centered view of enzyme function: residues do not contribute to catalysis only as isolated sequence positions, but as parts of an enzyme-substrate network that concentrates functional sensitivity around specific interaction regions.

A second implication is that mutation position alone is insufficient to explain mutational direction. ITS identified enriched mutation-sensitive regions, but higher-order topological features were required to distinguish activity-enhancing from lethal substitutions. Lethal mutations were associated with increased 1D Peak Count, consistent with fragmented or dispersed catalytic attention. By contrast, activity-enhancing mutations retained stronger 2D coordination, reflected by higher SVD Dominance, Global Gini index and Max Coupling. These patterns suggest that productive enzyme variants preserve sparse, globally organized residue-atom coupling, whereas lethal variants disrupt the coordinated topology needed for efficient turnover.

This interpretation extends previous deep-learning approaches for enzyme kinetics, which have mainly emphasized numerical prediction of $k_{cat}$ values. By combining sequence-derived protein representations, molecular graph features and bidirectional cross-attention, Interkcat provides a route from prediction to mechanistic interpretation. The framework also complements enzyme engineering strategies that seek to identify mutable positions while preserving catalytic function. In practice, ITS-defined regions may help prioritize residues with high functional sensitivity, whereas multidimensional topological features may help separate mutations that preserve coordinated catalysis from those that destabilize it.

Several limitations should be considered. First, attention-derived topology should not be interpreted as a direct physical contact map without structural or biochemical validation. The ITS and higher-order features provide model-derived hypotheses about interaction organization, not atomic-resolution mechanisms or direct measurements of catalytic propensity. Second, the mutation analysis depends on the availability and quality of curated kinetic measurements, which remain sparse and heterogeneous across enzyme families. Third, the wild-type-anchored grouping strategy controls substrate, organism and EC context, but this strict design reduces coverage and may underrepresent enzymes with limited mutant data. Finally, the activity-enhancing and lethal classes were defined using large $k_{cat}$ shifts, so the conclusions are strongest for mutations with substantial functional effects.

Despite these boundaries, the results support a coherent mechanistic model: catalytic efficiency and mutational outcome are encoded in the topology of enzyme-substrate coordination. Top-ITS zones identify where mutation-sensitive sites are statistically enriched, whereas higher-order attention features describe whether the interaction network remains coordinated or becomes fragmented. This distinction provides a practical framework for interpreting mutational effects and for guiding enzyme engineering toward substitutions that alter activity without collapsing catalytic organization.

## Methods

### Protein Sequence Representation via Partially Unfrozen ESM-2

To extract evolutionary and structural features from 1D enzyme sequences, we employed the pre-trained protein language model ESM-2 (esm2_t30_150M_UR50D). Amino acid sequences were tokenized, truncated to a maximum of 1024 tokens, and dynamically padded for uniform batch processing. Sequence embeddings were extracted from the model's last hidden state, with special boundary tokens (<cls> and <eos>) excluded to ensure exact spatial alignment with the amino acid residues in the subsequent cross-attention module.

To adapt these representations for continuous regression without overfitting the 150-million-parameter model or limiting task-specific expressivity, we applied a partial fine-tuning strategy. The majority of the base transformer layers were frozen to retain the pre-learned protein grammar, while the final two transformer layers and the contact head were selectively unfrozen. This enabled the terminal layers to adaptively refine the representations, generating contextualized residue-level embeddings ($E_p$) to drive the downstream bidirectional cross-attention mechanism.

**Molecular Graph Representation via Graph Neural Networks**

To capture the 2D topological structures and physicochemical properties of the catalytic substrates, we developed a customized GNN framework. Substrate SMILES strings were parsed into molecular graphs using RDKit[36] and PyG[37, 38], where nodes represent heavy atoms (with hydrogens removed to focus on the core molecular skeleton) and edges represent chemical bonds. To accommodate variable molecular sizes within computational batches, dynamic atomic masks were generated, with an empirical upper limit set to 100 nodes per molecule.

Atomic and bond features were projected into a high-dimensional hidden space using encoders and processed through a three-layer GNN architecture. Depending on the configuration, the network utilizes either standard GCN[39] or GAT[40]. During the forward message-passing phase, each atom iteratively updates its state by aggregating latent information from its local neighborhood. This aggregation incorporates edge attributes (bond features) to preserve chemical connectivity and stereoelectronic context.

To ensure stable model convergence and mitigate the vanishing gradient problem, each GNN layer is equipped with residual connections, Batch Normalization (BatchNorm1d), and ReLU activation functions. Following the graph convolutions, the updated node features are refined through a Multi-Layer Perceptron (Post-MLP) coupled with dropout regularization (rate = 0.1). This pipeline generates the contextualized atom-level molecular embeddings ($E_m$) alongside their valid-atom masks, which are fed into the downstream bidirectional cross-attention module to compute spatial-functional interactions with the enzyme sequences.

**Mathematical Formulation of the Bidirectional Cross-Attention Mechanism**

To capture the reciprocal interplay between the enzyme and its substrate, we implemented a bidirectional cross-attention module within the Interkcat framework. The formulation is based on the scaled dot-product attention mechanism[41] but is specifically configured to process two heterogeneous modalities.

Let $E_p \in \mathbb{R}^{B \times N \times D_p}$ denote the 1D protein embeddings extracted by the partially unfrozen ESM-2 module, and $E_m \in \mathbb{R}^{B \times M \times D_m}$ denote the atom-level molecular embeddings generated by the GNN, where $B$ is the batch size, and $N$ and $M$ are the sequence length and the number of atoms, respectively. To enable cross-modal computation, both embeddings are first mapped into a shared latent space of dimension $d$ (where $d = 256$) using linear projection matrices.

The bidirectional cross-attention establishes two distinct information flows: Protein → Molecule (p2m) and Molecule → Protein (m2p).

1. *Protein → Molecule Attention*

In this direction, the protein sequence acts as the Query ($Q_p$), actively scanning the substrate's atomic structure, which serves as the Key ($K_m$) and Value ($V_m$). The linear projections are formulated as:

$$Q_p = E_p W_{Q_p}, K_m = E_m W_{K_m}, V_m = E_m W_{V_m} \tag{1}$$

where $W_{Q_p} \in \mathbb{R}^{D_p \times d}$ and $W_{K_m}, W_{V_m} \in \mathbb{R}^{D_m \times d}$ are learnable weight matrices.

The attention logits are computed via a scaled dot-product and masked to negate padding tokens by adding a mask matrix $\mathcal{M}_m$ (where padding positions are set to $-\infty$). The localized attention weights $A_{p2m} \in \mathbb{R}^{B \times N \times M}$ and the enhanced protein representation $\tilde{E}_p$ are derived as follows:

$$A_{p2m} = \text{Softmax}\left(\frac{Q_p K_m^T}{\sqrt{d}} + \mathcal{M}_m\right) \tag{2}$$

$$\tilde{E}_p = A_{p2m} V_m \tag{3}$$

The $A_{p2m}$ matrix quantifies how each amino acid residue perceives the diverse reactive moieties of the substrate. It provides a global receptive field, ensuring that the enzyme's structural backbone is dynamically context-aware of the specific molecule it is catalyzing.

2. *Molecule → Protein Attention*

Conversely, the m2p flow allows the substrate atoms to act as the Query ( $Q_m$ ), interrogating the protein sequence ($K_p, V_p$) to pinpoint crucial functional sites:

$$Q_m = E_m W_{Q_m}, K_p = E_p W_{K_p}, V_p = E_p W_{V_p} \tag{4}$$

$$A_{m2p} = \text{Softmax}\left(\frac{Q_m K_p^T}{\sqrt{d}} + \mathcal{M}_p\right) \tag{5}$$

$$\tilde{E}_m = A_{m2p} V_p \tag{6}$$

This topological direction is inherently sparse and localized. By allowing individual substrate atoms to "vote" on the protein sequence, the $A_{m2p}$ matrix precisely identifies high-scoring topology regions.

3. *Integration and Prediction Head*

The resulting inter-coordinated representations ($\tilde{E_p}$ and $\tilde{E_m}$) encapsulate not only the isolated features of the entities but also their mutual spatial-functional interplay. Following pooling operations, these enhanced embeddings are concatenated to form a unified vector of dimension $2d$. To prevent overfitting and ensure stable convergence, we employed a carefully designed prediction head. It consists of projection matrices, Layer Normalization (LayerNorm), and dropout regularization ($p = 0.1$), ultimately decoded by a Multi-Layer Perceptron to output the continuous $k_{cat}$ value. Crucially, the raw attention matrices $A_{p2m}$ and $A_{m2p}$ computed in this module are preserved during the forward pass.

**Formulation of the Interaction Topology Score**

To quantitatively bridge the numerical attention weights with mechanistic interpretability, we developed ITS, a 1D sequence-level metric derived from the bidirectional cross-attention matrices. First, the raw m2p attention matrix is aggregated along the atomic axis and scaled via min-max normalization to yield a 1D weight vector, $W_{norm} \in [0,1]$. This vector represents the overall targeting intensity each amino acid receives from the substrate.

Concurrently, we evaluate the interaction specificity of each residue by calculating its normalized Shannon's Information Entropy ($H_{norm}$) based on the probability distribution extracted from the p2m attention matrix. A lower $H_{norm}$ value signifies highly concentrated and specific targeting, whereas a higher value indicates diffuse, non-specific contacts. To synergize these two complementary dimensions, ITS for each amino acid is formulated as an attention-derived interaction-topology metric.

$$ITS_i = W_{norm,i} \times (1 - H_{norm,i}) \tag{7}$$

By combining the normalized targeting intensity ($W_{norm}$) with interaction specificity ($1 - H_{norm}$), this formulation penalizes intense yet non-specific structural contacts. Consequently, a high ITS highlights residues that are both heavily targeted by the substrate and highly localized in their attention, providing a topology-aware landscape for testing enrichment of mutation-sensitive sites.

**Top-ITS Zone Formulation and Mutational Enrichment Analysis**

To evaluate whether ITS identifies sequence regions enriched for mutation-sensitive sites, we constructed one-dimensional (1D) topological profiles for a validation cohort consisting of the top 10 lethal and top 10 active single-point mutants. These variants represent the most extreme negative and positive kinetic shifts relative to their respective wild-type anchors.

To assess the spatial correlation between actual empirical mutations and the attention-derived topology profiles, the top five residues with the highest calculated ITS in a given sequence were defined as high-scoring topology centers. A sequence-based spatial tolerance radius of $\leq 5$ amino acids was applied upstream and downstream around these centers to account for localized structural propagation. Overlapping radii from proximal centers were algorithmically merged into continuous sequence windows, defined as Hit Zones.

To rigorously validate whether functional mutations preferentially cluster within these predicted landscapes, a statistical enrichment analysis was performed. For each enzyme sequence, the baseline probability of a random selection mechanism ($P_{\text{random}}$) was mathematically defined as the exact ratio of the aggregated Hit Zone length ($L_{\text{zone}}$) to the entire protein sequence length ($L_{\text{protein}}$):

$$P_{\text{random}} = \frac{L_{\text{zone}}}{L_{\text{protein}}} \quad (8)$$

Empirical mutation positions were then mapped onto these topological profiles. A successful "Hit" was recorded if an actual mutation fell within the boundaries of a merged Hit Zone. Finally, a one-sided binomial test was conducted to evaluate the statistical significance ($P$-value) of the observed empirical hit rate against the random background probability ($P_{\text{random}}$), accompanied by the quantification of distance error distributions from unhit positions to the nearest Hit Zone boundary.

**Mathematical Definition of Mechanistic Features for Mutation Analysis**

To systematically differentiate the structural and functional consequences of active versus lethal mutations, we extracted four higher-order topological features from the model's internal attention mechanisms. Instead of analyzing the molecule-to-protein ($A_{m2p} \in \mathbb{R}^{M\times N}$) and protein-to-molecule ($A_{p2m} \in \mathbb{R}^{N\times M}$) attention matrices in isolation, we first established a 2D Mutual Interaction Matrix ($M_{\text{mutual}} \in \mathbb{R}^{N\times M}$). This matrix was computed via the Hadamard (element-wise) product of the p2m matrix and the transposed m2p matrix:

$$M_{\text{mutual}} = A_{p2m} \odot A_{m2p}^{T} \quad (9)$$

This operation effectively filters out unidirectional noise, isolating highly confident reciprocal structural contacts between the $N$ amino acid residues and the $M$ substrate atoms. Based on the 1D ITS profile and the 2D mutual interaction matrix, the following four mechanistic features were defined:

1. *1D Peak Count (Spatial Fragmentation)*

To quantify the distribution of high-scoring topological regions along the linear protein sequence, we applied a topological peak-finding algorithm[42] to the 1D ITS profile. A localized ITS maximum was classified as a valid peak only if its prominence exceeded a 20% threshold of the sequence's global maximum ITS ($\text{Prominence} \geq 0.20 \times max(\text{ITS})$). The total number of identified peaks (Peak Count) serves as a proxy for the spatial integrity of the interaction-topology profile. A significantly elevated Peak Count implies that a mutation has caused fragmentation or erratic scattering of the enzyme's localized functional focus.

2. *Max Coupling Score (Interaction Intensity)*

The Max Coupling Score was defined as the absolute maximum value within the 2D mutual interaction matrix ($max(M_{\text{mutual}})$). This mathematical feature captures the single strongest reciprocal bond or structural coordination point between any specific amino acid residue and substrate atom, thereby reflecting the peak localized binding intensity preserved by the enzyme variant.

3. *Global Gini Index (Network Sparsity)*

To assess the inequality and sparsity of the overall interaction network, we calculated the Gini coefficient on the flattened 1D array of the $M_{\text{mutual}}$ matrix. For a flattened mutual vector $v$ of length $K$ (where $K = N \times M$), sorted in ascending order, the Gini index was computed as:

$$Gini = \frac{\sum_{i=1}^{K} (2i - K - 1)\, v_i}{K \sum_{i=1}^{K} v_i} \qquad (10)$$

A higher Global Gini index indicates a highly concentrated and sparsely distributed mutual attention network. Mechanistically, this means the enzyme efficiently focuses its catalytic energy on a few highly specific residue-atom pairs rather than dispersing it uniformly across non-specific structural contacts.

4. *SVD Dominance (Global Structural Coordination)*

To evaluate the concerted global structural coordination between the enzyme and the substrate, we performed Singular Value Decomposition (SVD) on the unflattened $M_{\text{mutual}}$ matrix. The SVD Dominance was defined as the ratio of the first (largest) singular value ($\sigma_1$) to the sum of all computed singular values:

$$\text{SVD Dominance} = \frac{\sigma_1}{\sum_i \sigma_i} \tag{11}$$

From a physical perspective, a high SVD Dominance score implies that the enzyme-substrate interplay is strongly governed by a single, highly coordinated principal interaction component. A severe drop in this score mathematically reflects the collapse of 2D structural coordination, which serves as a hallmark of lethal mutations.

**Sequence Alignment and Mutational Event Parsing**

To systematically map the precise locations of mutation sites within the previously defined groups, each mutant sequence was computationally aligned against its corresponding intra-group WT reference sequence. This pairwise alignment was executed using the Needleman-Wunsch global alignment algorithm, implemented via the PairwiseAligner module within the Biopython library[43, 44]. To ensure biologically meaningful alignments and appropriately restrict artifactual gaps, the scoring matrix was strictly configured with a match score of 2, a mismatch penalty of -1, a gap opening penalty of -10, and a gap extension penalty of -0.5. Following the alignment, a customized parsing algorithm was deployed to systematically extract the exact mutational events from the aligned sequence pairs. This parsing procedure systematically identified and categorized all sequence discrepancies, encompassing single or multiple amino acid substitutions, insertions, and deletions, while simultaneously registering their precise spatial coordinates along the primary WT sequence.

## Data availability

The data that support the findings of this study, including the EITLEM-Kinetics benchmark dataset and the pre-trained Interkcat models, are available in Code Ocean at https://codeocean.com/capsule/5279530/tree. Source data are provided with this paper.

## Code availability

The custom computer code is available on GitHub at https://github.com/ZWR0/Interkcat, and a persistent, reproducible version including data and pre-trained models is available in Code Ocean at https://codeocean.com/capsule/5279530/tree.

## Acknowledgments

The authors thank Prof. Akutsu for providing the computational equipment used for model training.

## Author contributions

W.Z. conceptualized the study and implemented the Interkcat deep learning model. T.T. supervised the project. Both authors co-wrote the original draft and contributed to the final manuscript revisions.

## Funding

This work was supported by JST SPRING [grant number JPMJSP2110] and JSPS KAKENHI [grant number 26K02972].

## Competing interests

The authors declare no competing interests.